\documentclass{emulateapj}
\usepackage{url}

\shorttitle{Ages of Four Clusters}
\shortauthors{Jeffery et al.}

\begin{document}

\title{A Bayesian Analysis of the Ages of Four Open Clusters}

\author{Elizabeth J. Jeffery}
  \affil{Department of Physics and Astronomy, Brigham Young University, 
    Provo, UT 84602, USA}
\email{ejeffery@byu.edu}
\author{Ted von Hippel}
  \affil{Center for Space an Atmospheric Research, Embry-Riddle Aeronautical University,
    Daytona Beach, FL 32114, USA}
\author{David A. van Dyk}
  \affil{Statistics Section, Imperial College London, London, UK}
\author{David C. Stenning}
  \affil{Sorbonne Universit\'{e}s, UPMC-CNRS, UMR 7095, Institut d'Astrophysique de Paris, F-75014 
    Paris, France}
\author{Elliot Robinson}
  \affil{Argiope Technical Solutions, Ft White, FL, USA}
\author{Nathan Stein}
  \affil{The Wharton School, University of Pennsylvania, 
    Philadelphia, PA 19104, USA}
\author{William H. Jefferys}
  \affil{University of Texas, Austin, TX, USA}
  \affil{University of Vermont, Burlington, VT, USA}

\begin{abstract}

   In this paper we apply a Bayesian technique to determine the best fit of stellar evolution models to find the main sequence turn off age and other cluster parameters of four intermediate-age open clusters: NGC 2360, NGC 2477, NGC 2660, and NGC 3960. Our algorithm utilizes a Markov chain Monte Carlo technique to fit these various parameters, objectively finding the best-fit isochrone for each cluster. The result is a high-precision isochrone fit. We compare these results with the those of traditional ``by-eye" isochrone fitting methods. By applying this Bayesian technique to NGC 2360, NGC 2477, NGC 2660, and NGC 3960, we determine the ages of these clusters to be 1.35 $\pm$ 0.05, 1.02 $\pm$ 0.02, 1.64 $\pm$ 0.04, and 0.860 $\pm$ 0.04 Gyr, respectively. The results of this paper continue our effort to determine cluster ages to higher precision than that offered by these traditional methods of isochrone fitting.

\end{abstract}


\section{Introduction}
   \label{intro}

   Star clusters have long been important tools for studying stellar evolution, specifically because they play {\it the} pivotal role in determining the ages of stars. The most commonly used method for measuring the age of an open star cluster involves fitting an isochrone to the cluster's observed color-magnitude diagram (CMD), specifically to the cluster's main sequence turn off (MSTO). Generating and fitting isochrones to a cluster CMD to determine its age also requires knowledge of the cluster's metallicity, distance, and reddening. Oftentimes, finding a best fit of these three parameters (plus age) is a subjective process, as some of these parameters are correlated with each other. This difficulty is reflected in isochrones that appear to fit the CMD equally well with various combinations of cluster parameters (see, for example, Figure 2 of VandenBerg \& Stetson 2004). Moreover, the fit of the MSTO can be challenging and isochrones may give inconsistent results in different filters, even when using the same cluster parameters (see, for example, Figure 10 of Sarajedini et al. 1999).

   An independent method to measure the age of a cluster involves using the cluster white dwarfs (WDs). Because a WD's luminosity is directly related to its cooling time (Mestel 1952; Winget et al. 1987), this information, along with WD masses and atmospheric types, provide the WD cooling age and ultimately the cluster age. Measuring and comparing the MSTO age and the WD age of a cluster is currently the best means to test and calibrate both methods and their underlying theory.

   We seek a more objective way to fit isochrones to cluster CMDs to more precisely determine ages from both the MSTO and the cluster WDs. High-precision ages will allow for more meaningful comparison and calibrations between the two methods. To this end, our group has developed and successfully implemented a robust technique that utilizes Bayesian statistical methods. The Bayesian method determines the posterior distribution of model parameters, resulting in something akin to a best fit. In this paper, we determine the age and cluster parameters of metallicity, distance, and reddening for four intermediate-age ($\sim$1 Gyr) open clusters: NGC 2360, NGC 2477, NGC 2660, and NGC 3960. 

   Our motivation in choosing these clusters is primarily related to testing WD models. Clusters in this age range are sensitive to crystallization and phase separation of carbon and oxygen in WDs. Our group has obtained deep observations of these clusters with the Hubble Space Telescope, and we will analyze the WD sequences in these clusters in a future companion paper. In this paper we focus on new photometric ground-based data we have obtained for the purpose of measuring the MSTO age and improving cluster parameters for these four clusters.

   We have organized this paper as follows: we discuss the clusters and the observations in Section \ref{obs}, including observed CMDs for the complete field around each cluster; in Section \ref{isochrones} we determine the MSTO age for each cluster using traditional methods of fitting isochrones, largely by eye; in Section \ref{bayes} we describe the Bayesian technique and how it is applied to each cluster (including the necessary prior distributions on several parameters);  we discuss the results in Section \ref{results}; and we end with concluding remarks in Section \ref{conclusion}.

\section{Observations and Photometry}
  \label{obs}

   The four clusters in this study (NGC 2360, NGC 2477, NGC 2660, and NGC 3960) have long histories of prior observations. We summarize some of the previous determinations of these clusters' parameters in Tables \ref{lit2360} \--- \ref{lit3960}. For consistency, values of distance and reddening are reported in the Tables as $(m - M)_{V}$ and $A_{V}$, regardless of how they are reported in the original source. If the original source reported $E(B - V)$, we converted this to $A_{V}$ using the relationship

\begin{equation}
    A_{V} = 3.1\ E(B-V). 
\end{equation}

\noindent
Similarly, when a literature source reports unreddened distance modulus $(m-M)_{0}$, we converted it to the apparent distance modulus using the standard definition:

\begin{equation}
    (m-M)_{V} = (m-M)_{0} + A_{V}.
\end{equation}

   For this study we obtained new observations of each of these clusters. In this section, we describe the observations, data reduction, and the process of obtaining photometry.


\begin{table}[t]
  \begin{center}
   \caption{Cluster parameters from the literature for NGC 2360}
      \label{lit2360}
  \begin{tabular}{lcccc}

    \hline
 Age (Gyr) & $(m-M)_{V}$ & $A_{V}$ & [Fe/H] & Reference \\
    \hline


  0.80  & \---  & \--- & \---  &  1 \\
  0.85  & \---  & \--- & $-$0.14 &  2 \\ 
  1.00  & 10.40 & 0.28 & \---  &  3 \\
  1.15  & 10.40 & 0.22 & +0.07 &  4 \\
  1.40  & 10.45 & 0.22 & \---  &  5 \\
  1.80 &  10.09 & 0.19 & \---  &  6 \\
  1.90  & 10.70 & 0.28 & $-$0.28 &  7 \\
  2.20  & 10.50 & 0.25 & +0.07 &  8 \\
  \---  & 10.35 & 0.28 & $-$0.15 &  9 \\
  \---  & 10.56 & 0.28 & $-$0.26 & 10 \\
  \---  & \---  & \--- & $-$0.16 & 11 \\
  \---  & \---  & \--- & $-$0.07 & 12 \\


   \hline
  \end{tabular}
   \end{center}
 \textbf{References.} (1) Patenaude 1978; (2) Salaris et al. 2004; (3) Meynet et al. 1993; (4) Hamdani, et al. 2000; (5) Mazzei \& Pigatto 1988; (6) Gunes, Karatas, \& Bonatto 2012; (7) Friel \& Janes 1993; (8) Mermilliod \& Mayor, 1990; (9) Twarog, et al. 1997; (10) Friel et al. 2002; (11) Claria et al. 2008; (12) Reddy, Giridhar, \& Lambert 2012

\end{table}



\begin{table}[t]
  \begin{center}
   \caption{Cluster parameters from the literature for NGC 2477}
      \label{lit2477}
  \begin{tabular}{lcccc}

    \hline
 Age (Gyr) & $(m-M)_{V}$ & $A_{V}^{a}$ & [Fe/H] &  Reference \\
    \hline


  1.0  & 11.43 & 0.93 & \---   &  1 \\
  1.0  & 11.45 & 0.713 & 0.00  &  2 \\
  1.0   &  \--- & \--- & \--- & 3 \\  
  1.04$^{b}$ & 11.4 & 0.60 & $-$0.1  &  4 \\
  1.0  & \---  & \--- & $-$0.14 &  5 \\
  1.3  & 11.60 & 0.93 & $-$0.05 & 6 \\
  1.5  & 11.48 & 0.868 & \---  &  7 \\


   \hline
  \end{tabular} 
   \end{center}
 \textbf{Notes.} \\
 $^{a}$Average value. \\
  $^{b}$White dwarf age
 \\ \textbf{References.} (1) Kassis et al. 1997; (2) Salaris et al. 2004; (3) von Hippel, Gilmore, \& Jones 1995; (4) Jeffery et al. 2011; (5) Eigenbrod et al. 2004; (6) Friel \& Janes 1993; (7) Hartwick, et al. 1972.

\end{table}



\begin{table}[t]
  \begin{center}
   \caption{Cluster parameters from the literature for NGC 2660}
      \label{lit2660}
  \begin{tabular}{lcccc}

    \hline
 Age (Gyr) & $(m-M)_{V}$ & $A_{V}$ & [Fe/H] &  Reference \\
    \hline


  0.73 & \---  & \--- & $-$0.55 &  1 \\
  0.95 & \---  & 1.33 & +0.04 &  2 \\
  1.0  & 13.44 & 1.24 & 0.00  &  3 \\
  1.0  & 13.44 & 1.24 & +0.02 &  4 \\
  1.1  & 13.94 & 1.24 & \---  &  5 \\
  1.2  & 13.48 & 1.18 & +0.103$^{a}$ &  6 \\
  1.7  &  \--- & 1.15 & $-$1.05 &  7 \\


   \hline
  \end{tabular}
 \end{center}
 \textbf{Note.}
  $^{a}$The paper cited reports NGC 2660 as having the metallicity of the Hyades.  The value in the table is the metallicity of the Hyades from Taylor \& Joner 2005. \\
 \textbf{References.} (1) Salaris et al. 2004; (2) Bragaglia et al. 2008; (3) Sandrelli et al. 1999; (4) Sestito et al. 2006; (5) Mazzei \& Pigatto 1988; (6) Hartwick \& Hesser 1973; (7)  Geisler, Claria, \& Minniti 1992.

\end{table}



\begin{table}[t]
  \begin{center}
   \caption{Cluster parameters from the literature for NGC 3960}
      \label{lit3960}
  \begin{tabular}{lcccc}

    \hline
 Age (Gyr) & $(m-M)_{V}$ & $A_{V}$ & [Fe/H] &  Reference \\
    \hline


  0.6           & \---  & \---          & \---  &  1 \\
  0.625$^{a}$   & 12.0  & 0.899         & $-$0.30 &  2 \\
  0.9           & 11.6  & 0.899$^{b}$   & $-$0.12 &  3 \\
  0.9\---1.4   & 12.25 & 0.899$^{c}$   & \---  &  4 \\
  1.1           & 11.60 & 0.403         & \---  &  5 \\
  1.0           & 12.0  & 0.899         & $-$0.34 &  6 \\
  1.0           & 11.60 & 0.899         & +0.04 &  7 \\
 \---           & 12.15 & 0.961         & $-$0.17 &  8 \\
 \---           & \---  & \---          & $-$0.04 &  9 \\


   \hline
  \end{tabular}
   \end{center}
 \textbf{References.} (1) Carraro et al. 1998; (2) Janes 1981; (3) Bragaglia, et al. 2006; (4) Prisinzano et al. 2004; (5) Bonatto \& Bica 2006; (6) Friel \& Janes 1993; (7) Sestito, et al. 2006; (8) Twarog et al. 1997; (9) Heiter et al. 2014; $^{a}$The paper cited reports NGC 3960 as having the age of the Hyades. The value in the table is the age of the Hyades from Perryman et al. (1998). $^{b}$Average value; $^{c}$Value in the cluster center.

\end{table}


   \subsection{Observations}
   \label{ctio_obs}

   We observed these four clusters using the Y4KCam CCD on the 1.0m telescope at Cerro Tololo Inter-American Observatory; this telescope is operated by the Small and Moderate Research Telescope System (SMARTS) consortium.\footnote{\url{http://www.astro.yale.edu/smarts/}} The Y4KCam CCD has a 4064 $\times$ 4064 chip with a plate scale of 0.298 arcseconds per pixel, giving it a field of view (FOV) of 20' $\times$ 20', ideal for cluster 
observations.  The data discussed here were taken over the course of three nights, in standard $BVI$ filters. We present a log of observations in Table \ref{obs_table}. In addition to cluster observations, we observed Landolt (1992) standard stars to transform the data to the standard system.

\begin{table}[t]
  \begin{center}
   \caption{Log of observations}
      \label{obs_table}
  \begin{tabular}{cccrcc}

    \hline
  &   &   &  Exposure & No. of \\
  \multicolumn{1}{c}{Cluster} & \multicolumn{1}{c}{Date (UT)} & \multicolumn{1}{c}{Filters} & Time (s) & Exposures  \\
    \hline

 NGC 2360 & 2007 Apr 27 & $BVI$ &    10/10/10 & 3 \\
          &      $''$    & $BVI$ &   120/60/50 & 3 \\
          &     $''$     & $BVI$ & 600/300/400 & 3 \\
 NGC 2477 & 2007 Apr 26 & $BVI$ &    10/10/10 & 3 \\
          &      $''$    & $BVI$ &   120/60/50 & 3 \\
          &     $''$     & $BVI$ & 600/300/400 & 3 \\
 NGC 2660 & 2007 Apr 25 & $BVI$ &   120/60/50 & 3 \\
 NGC 3960 & 2007 Apr 27 & $BVI$ &    10/10/10 & 3 \\
          &      $''$    & $BVI$ &   120/60/50 & 3 \\
          &     $''$     & $BVI$ & 600/300/400 & 3 \\

   \hline
  \end{tabular}
 \end{center}
\end{table}


     We reduced the raw data frames (including bias correction and flat-fielding, etc.) using a pipeline\footnote{\url{http://www.lowell.edu/users/massey/obins/y4kcamred.html}} developed and kindly  provided by Phillip Massey, which runs in IRAF.\footnote{IRAF is distributed by the National Optical  Astronomy Observatories, which are operated by the Association of Universities for Research in Astronomy, Inc., under cooperative agreement with the National Science Foundation.} We note that during our run, the northwest quadrant of the CCD was non-functioning, so the actual FOV available was 75\% its usual value. The reduction script assumes four working quadrants, and though only three were functioning during our run, we still executed the script as normal, treating the dead quadrant as if it were functioning, then disregarding it in the end. This did not affect the reductions
of the other quadrants.

   \subsection{Photometry of Cluster Images}
   \label{phot}

     For all cluster images, source finding on the individual, science-ready images was done using the source finding routine SExtractor (Bertin \& Arnouts 1996). We measured instrumental magnitudes by utilizing aperture photometry routines in the IRAF APPHOT package. To transform instrumental magnitudes to the standard system, we applied the following transformation equations:

\begin{equation}\label{phot_trans_B}
      b = B + b_{0} + b_{1}X + b_{2}(B - V) \\
\end{equation}

\begin{equation}\label{phot_trans_V}
      v = V + v_{0} + v_{1}X + v_{2}(B - V) \\
\end{equation}

\begin{equation}\label{phot_trans_I}
      i = I + i_{0} + i_{1}X + i_{2}(V - I). \\
\end{equation}

\noindent
Variables in these equations are defined in the standard way: small letters are used to represent instrumental magnitudes, while uppercase letters are standardized magnitudes; $x_{0}$ is the zero point for a given filter; $x_{1}$ is the extinction coefficient applied to an observation taken at an airmass $X$; and $x_{2}$ is the color term. In Table \ref{trans_sol} we have listed these values for each night of our observations.

\begin{table}[t]
  \begin{center}
   \caption{Transformation Equation Coefficients}
      \label{trans_sol}
  \begin{tabular}{cccc}

    \hline

              &          2007 Apr 25          &           2007 Apr 26          &         2007 Apr 27          \\
\hline
 $b_{0}$  &  $-$22.871 $\pm$ 0.040  &  $-$22.731 $\pm$ 0.051  &  $-$22.774 $\pm$ 0.115 \\
 $b_{1}$  &  0.547 $\pm$ 0.023        &  0.409 $\pm$ 0.029         & 0.486 $\pm$ 0.087         \\
 $b_{2}$  &  0.091 $\pm$ 0.020        &  0.107 $\pm$ 0.024         & 0.096 $\pm$ 0.025         \\
\hline
 $v_{0}$  &  $-$22.934 $\pm$ 0.042  & $-$22.978 $\pm$ 0.024   & $-$22.984 $\pm$ 0.084  \\  
 $v_{1}$  &  0.292 $\pm$ 0.024         & 0.267 $\pm$ 0.014         &  0.348 $\pm$ 0.064        \\
 $v_{2}$  &  $-$0.086 $\pm$ 0.019    & $-$0.066 $\pm$ 0.010    & $-$0.098 $\pm$ 0.020     \\
\hline
 $i_{0}$  &  $-$22.048 $\pm$ 0.043  &  $-$21.882 $\pm$ 0.042   & $-$22.168 $\pm$ 0.071  \\
 $i_{1}$  &  0.170 $\pm$ 0.027         &  0.032 $\pm$ 0.025         & 0.245 $\pm$ 0.053        \\
 $i_{2}$  &  $-$0.050 $\pm$ 0.017    & $-$0.052 $\pm$ 0.015      & $-$0.036 $\pm$ 0.012   \\

   \hline
  \end{tabular}
 \end{center}
\end{table}

    We determined the coefficients of the transformation equations using Landolt (1992) standards, utilizing the IRAF/PHOTCAL package. Once these coefficients were determined, we applied them  to the instrumental magnitudes of our program stars to transform them to the standard system. Multiple observations of the same star were then averaged together to obtain magnitude, color, and error values for each star.

\begin{figure*}
\begin{center}
  \epsscale{0.9}
     \plotone{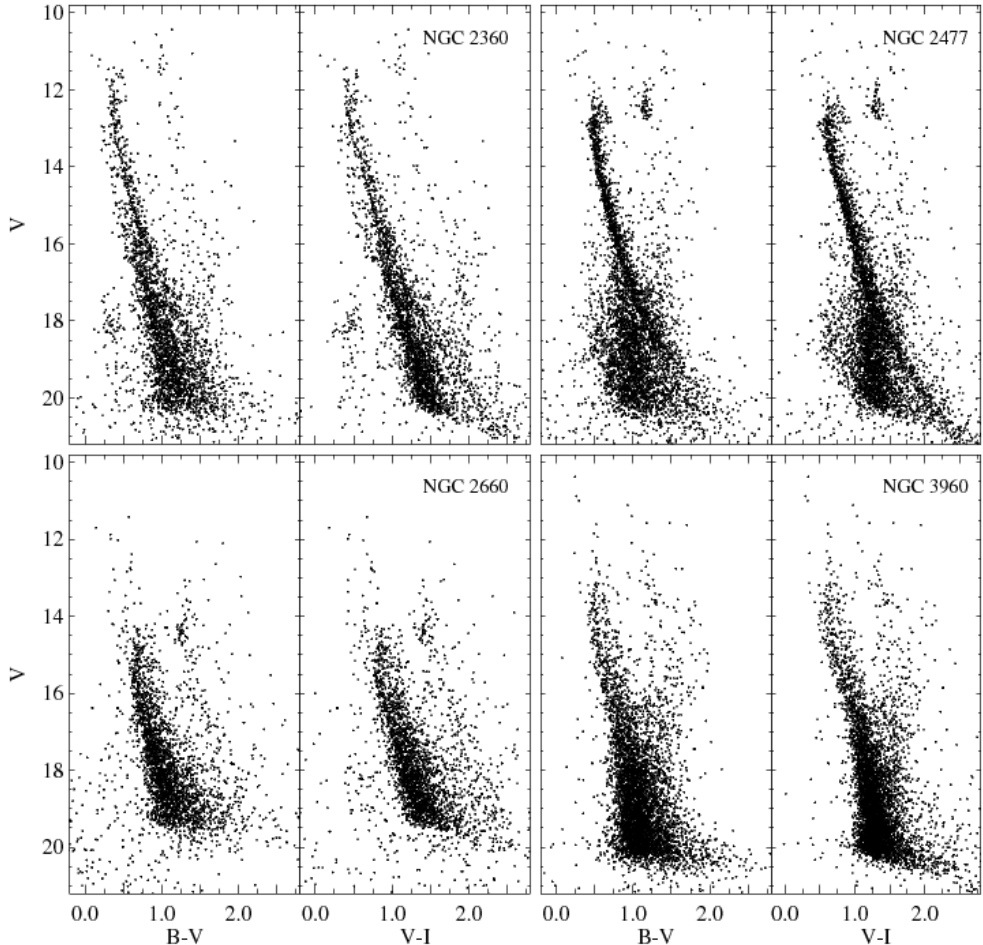}
         \caption{The $B - V$ and $V - I$ CMDs for the complete fields 
	around each of the four clusters in this study.}
\label{cmd_all}
\end{center}
\end{figure*}

     Once standard magnitudes and colors were determined for stars on the clusters images, we constructed CMDs for each cluster. We present the complete CMDs of the fields of all four clusters in Figure \ref{cmd_all}. In each case, the left panel is the CMD in the $B-V$ color, and the right panel is the $V-I$ color. Clearly visible in each CMD is the cluster MS as well as the MSTO and giant stars, along with an abundance of field stars.

\section{Fitting Isochrones}
  \label{isochrones}

     With CMDs of the clusters, we are able to determine the MSTO age of each. In this section, we apply this technique in the ``classical" way, determining a best-fit isochrone by eye. These are compared with objective fits obtained via a Bayesian algorithm in Section  \ref{bayes}. For fitting MS isochrones, we chose to use the Dartmouth Stellar Evolution Database (Dotter et al. 2008). To create the isochrones, we utilized the online form.\footnote{\url{http://stellar.dartmouth.edu/models/isolf\_new.html}} Our procedure was to fix metallicity, guided by spectroscopically derived [Fe/H] values reported in the literature, start with a distance modulus and absorption consistent with values reported in the literature, and then iteratively adjust age, $(m-M)_{V}$ and $A_{V}$, until we achieved a best fit in both CMDs, as judged by eye.

\begin{figure*}
\begin{center}
  \epsscale{0.9}
     \plotone{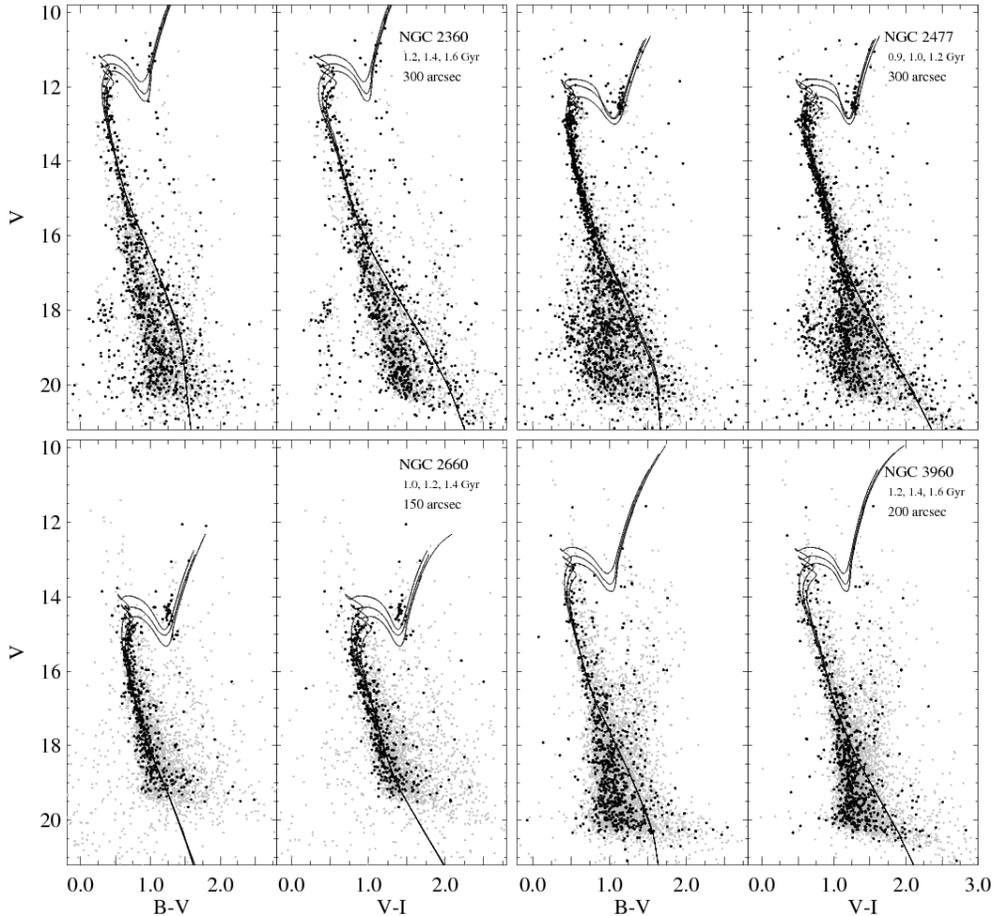}
         \caption{The $B - V$ and $V - I$ CMDs with isochrones 
        overlaid. Gray points represent all stars within the observed field of 
	view and black points represent those objects 
	within a certain radius from the cluster core, as indicated 
	in the figure. Cluster parameters used for 
	fitting are listed in Table \ref{iso_params}.}
\label{cmd_iso}
\end{center}
\end{figure*}

   To alleviate contamination from likely field stars when fitting isochrones, and thus improve confidence in the fit to the MSTO, only stars within a certain radius of the approximate cluster center were used. (This radius is specified for each cluster in Figure \ref{cmd_iso}.) In the CMDs in Figure \ref{cmd_iso}, the black points are within this radius and the gray points are outside this radius. Although this method is crude for identifying likely cluster members, it is an adequate first attempt at cleaning field star contamination and is sufficient for our purposes.

   In Figure \ref{cmd_iso} we display our results of fitting isochrones to both $B-V$ and $V-I$ CMDs. In each case we have included isochrones for three different ages: a best fit along with two isochrones that bracket the MSTO, giving an estimate of the uncertainty of the age. Uncertainty in the fit of the isochrone due to the spread in the MSTO region can be caused by, e.g., unresolved binaries or photometric uncertainty. Using this method of fitting isochrones by eye, the reported age is  the middle isochrone (as the best fit), and the uncertainty is estimated from the upper and lower isochrones, giving an upper and lower bound to the age. We do not include a more rigorous error analysis at this point because our Bayesian method provides principled error estimates automatically (Section \ref{bayes}). We have included the best-fit age value for each cluster in Table \ref{iso_params}. 

   The isochrones shown in Figure \ref{cmd_iso} were generated using metallicity values consistent with literature values. We have also shifted the isochrones appropriately for distance and reddening and we present these best-fit values, including the metallicity used, in Table \ref{iso_params}.

\begin{table}[t]
  \begin{center}
   \caption{``By-eye" best-fit values of cluster parameters}
      \label{iso_params}
  \begin{tabular}{ccccc}

    \hline
 Cluster & Age (Gyr) &  [Fe/H] & $(m - M)_{V}$ & $A_{V}$ \\
    \hline


 NGC 2360 &  1.4 $\pm$ 0.2 & $-$0.20 & 10.25 & 0.25 \\
 NGC 2477 &  1.0 $\pm$ 0.2 & $-$0.10 & 11.35 & 0.75 \\
 NGC 2660 &  1.2 $\pm$ 0.2 & 0.00    & 13.35 & 1.05 \\
 NGC 3960 &  1.4 $\pm$ 0.2 & $-$0.30 & 11.80 & 0.65 \\

\hline

 \multicolumn{5}{l}{Note: The metallicity values used to generate each isochrone} \\
 \multicolumn{3}{l}{were guided by literature values.} & & \\

   \end{tabular}
 \end{center}
\end{table}


   One of the severe limitations of this technique is that one must estimate the best-fit isochrone by 
simultaneously fitting multiple parameters in multiple CMDs. Very little can be done to robustly 
determine error values of the fit, especially in distance and reddening. 

\section{A Bayesian Approach}
  \label{bayes}
     
     Although the classical technique of determining cluster ages presented above has been used for decades, modern computational and statistical techniques allow for more principled, robust, and reliable fitting. To this end, we have developed a sophisticated software suite to objectively fit models to our data, utilizing Bayesian statistics: Bayesian Analysis for Stellar Evolution with Nine Parameters (BASE-9). The BASE-9 source code is freely available for download on Github,\footnote{\url{https://github.com/argiopetech/base}, accessed 2016 May 20} or the executables can be accessed via Amazon Web Services. The use of BASE-9 is described in detail by von Hippel et al. (2014).

   \subsection{Overview of the Technique}
    \label{overview}

      A more in depth description of the Bayesian technique (including the explicit mathematical equations for the likelihood) employed here can be found in previous papers published by our group (e.g., von Hippel et al. 2006; DeGennaro et al. 2009; van Dyk et al. 2009; Stein et al. 2013). Briefly, BASE-9 derives posterior distributions for various cluster and stellar parameters by utilizing Bayesian analysis methods. Because of the high dimensionality and complex nature of these distributions, we utilize an adaptive Markov chain Monte Carlo (MCMC) technique to sample the joint posterior distributions of the different parameters (Stenning et al. 2016). BASE-9 uses the stellar evolution model to generate theoretical photometry values of cluster stars, and compares them to the observed photometry, including photometric errors, to produce the parameter values at each step. Each step in the MCMC chain consists of a set of cluster parameters, namely age, metallicity, distance, and reddening. The convergent MCMC chain provides a sample from the posterior distribution of cluster parameters, and can be used to compute means and intervals as parameter estimates and uncertainties.

    BASE-9 is capable of estimating the posterior probability distributions for six cluster-wide parameters, and three individual stellar properties (nine total). The cluster properties are age, metallicity, distance modulus, line-of-sight absorption ($A_{V}$), helium abundance, and the initial-final mass relation (IFMR); individual stellar properties that can be estimated are primary mass, and the ratio of secondary to primary mass (if a binary system). The model also accounts for field star contamination, and we can use that to compute the probability that a given star is a cluster member. In our current analysis we only analyze four cluster-wide parameters (age, metallicity, distance modulus, and absorption). We treat helium abundance and IFMR as fixed quantities, and marginalize over stellar masses. We assign each star the same prior membership probability, and in each step of the MCMC chain marginalize over cluster membership status.

   A detailed description of the field star modeling process can be found in Stein et al. (2013). To summarize, for each star we introduce an additional indicator variable, $Z_{j}$, that is equal to one if star $j$ is a cluster star and is equal to zero otherwise. This allows us to specify separate statistical models for the observed photometric magnitudes of cluster stars versus those of field stars. We use a Gaussian model for the photometric magnitudes of cluster stars, with known (independent) measurement errors contained in the (diagonal) variance-covariance matrix. For field stars, we use a simple model whereby the magnitudes are assumed to be uniformly distributed over the range of the data; this simple model is adequate for identifying field stars (see Stenning et. al 2016 for a simulation study). The final statistical model for a star can then be expressed as $Z_{j}\times[\textrm{Cluster Star Model}] + (1-Z_{j})\times[\textrm{Field Star Model}]$. Such models are known as {\it finite mixture distributions}, and represent the fact that the observed data contains a mixture of two subgroups: cluster stars and field stars. A key advantage is that we do not have to specify a priori which stars are cluster members and which are field stars.

   One of the advantages of a Bayesian analysis is that it offers a principled method for combining information from multiple sources in a single coherent analysis. Typically, information external to the current data is summarized in the prior distribution and when combined with information in the data yields the posterior distribution. Thus, the posterior is a complete statistical summary of information from both sources for the parameters and can be used to derive parameter estimates and error bars. 

     The precision of the parameters discussed in the following sections is internal precision, rather than external accuracy.  Our technique objectively determines the posterior distribution of model parameters fit to the data, with the center of that distribution representing something like a best fit; it cannot assess the physical accuracy of the model itself.  

   We again use the MS evolution timescale models of Dotter et al. (2008). We note that other model sets can be employed within BASE-9 as well.

     \subsection{Input Data}
     \label{bayes_prep_msto}
     
\begin{figure*}
\begin{center}
  \epsscale{0.8}
     \plotone{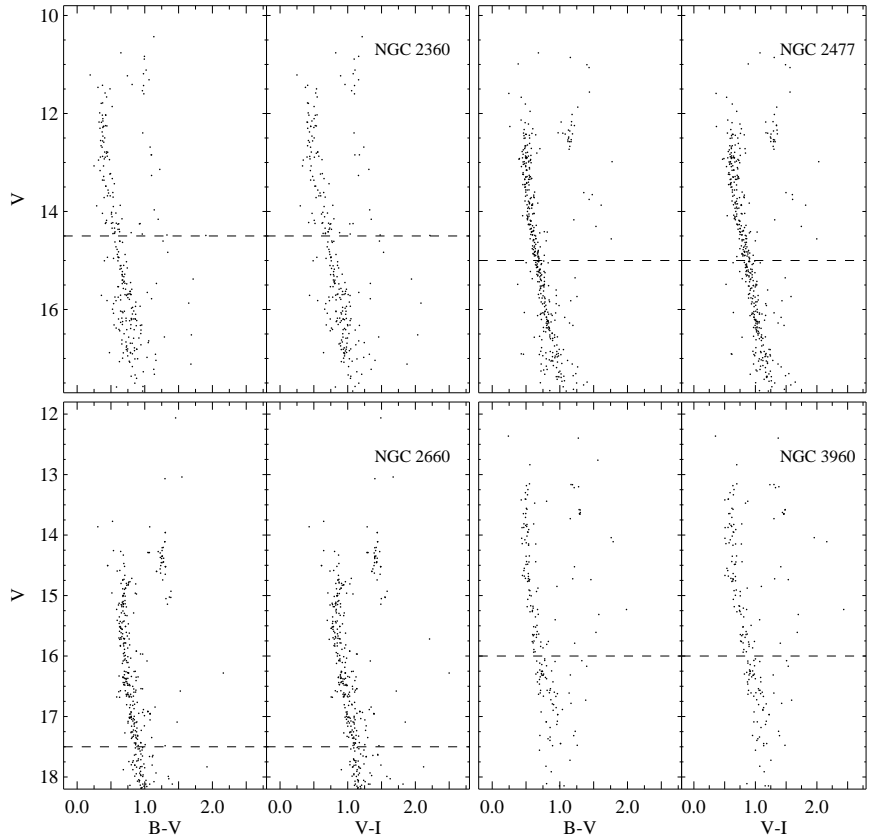}
         \caption{The $B - V$ and $V - I$ CMDs of the data that were input 
	to BASE-9. The horizontal dashed line for each cluster indicates the cutoff value, 
	below which stars were not included in the analysis.}
\label{cmd_bayes}
\end{center}
\end{figure*}

   In preparation for running our cluster data through BASE-9, we first culled the complete photometry list to include only stars with magnitude and color errors less than 0.1 mag. Additionally, to alleviate confusion caused by the high number of field stars, we have excluded stars that are further than a particular radius from the approximate cluster center. (We use the same radii as previously discussed; see Figure \ref{cmd_iso}.) We display in Figure \ref{cmd_bayes} the $B-V$ and $V-I$ CMDs for the stars included in the Bayesian analyses. 

   The horizontal dashed lines in Figure \ref{cmd_bayes} indicate the imposed $V$ magnitude cutoff for each cluster. Stars fainter than the cutoff were excluded from our analysis. Our primary motivation for doing this is to avoid fitting the lower MS. While models tend to fit the upper MS well, most do a poor job at fitting the lower MS. Limitations of the isochrones affect the fitted results, as BASE-9 cannot assess the physical reliability of the model (this specific issue was explored extensively by DeGennaro et al. 2009). This $V$ magnitude cutoff was chosen to be approximately 3 magnitudes below the turnoff of each cluster. This choice was guided by results from DeGennaro et al. (2009).

   The Bayesian technique requires prior distributions for all parameters. Prior distributions on metallicity, distance modulus, and reddening were assumed to be Gaussian (see Table \ref{priors_table}), and were determined using the mean and standard deviation of literature values (see Tables \ref{lit2360} \-- \ref{lit3960}). The prior distribution on reddening is truncated at zero because $A_{V}$ is always positive. We used an uninformative prior for cluster age that was flat in log(age), truncated to the realistic range of 0.25 Gyr to 15 Gyr.


\begin{table}[t!]
  \begin{center}
   \caption{Prior distributions used for each cluster}
      \label{priors_table}
  \begin{tabular}{cccc}

    \hline
%
 Cluster &   [Fe/H]  & $(m - M)_{V}$ & $A_{V}$ \\
    \hline


NGC 2360 &     0.00 $\pm 0.15$ & 10.50 $\pm 0.15$ & 0.25 $\pm 0.04$ \\
NGC 2477 &  $-$0.10 $\pm 0.10$ & 11.50 $\pm 0.10$ & 0.85 $\pm 0.10$ \\
NGC 2660 &  $-$0.20 $\pm 0.40$ & 13.50 $\pm 0.20$ & 1.23 $\pm 0.06$ \\
NGC 3960 &  $-$0.15 $\pm 0.15$ & 11.90 $\pm 0.25$ & 0.91 $\pm 0.03$ \\

\hline

   \end{tabular}
 \end{center}
\end{table}


   We ran BASE-9 on each cluster a total of twelve times, each time running the chain for 26,000 steps. For each cluster, the 12 runs were divided into 4 sets of 3 runs: each set used a particular set of starting values for metallicity, distance, and reddening, and 1 of 3 different age starting values. We list these starting values for each cluster in Table \ref{starting}. Changing the starting values allowed us to test the robustness of our algorithm in determining a consistent posterior distribution, regardless of the starting value of the MCMC chain.


\begin{table}[t]
  \begin{center}
   \caption{Starting values for BASE-9 convergence tests}
      \label{starting}
  \begin{tabular}{ccccc}

    \hline
 Cluster    &         &          &               &         \\ 
 (log(Age)) & Set \# & [Fe/H] & $(m - M)_{V}$ & $A_{V}$ \\ 
    \hline


 NGC 2360        & 1 &  0.0   & 10.50 & 0.25 \\ 
 (9.0, 9.1, 9.2) & 2 & $-$0.1 & 10.35 & 0.20 \\ 
                 & 3 & $-$0.2 & 10.70 & 0.30 \\
                 & 4 & +0.1   & 10.40 & 0.20 \\
 NGC 2477        & 1 &  0.0   & 11.45 & 0.90 \\
 (9.0, 9.1, 9.2) & 2 & $-$0.1 & 11.50 & 0.70 \\
                 & 3 & $-$0.2 & 11.60 & 0.80 \\
                 & 4 & +0.1   & 11.40 & 0.60 \\
 NGC 2660        & 1 & $-$0.2 & 13.50 & 1.20 \\
 (9.1, 9.2, 9.3) & 2 & $-$0.1 & 13.00 & 1.20 \\
                 & 3 & $-$0.05 & 13.90 & 1.30 \\
                 & 4 & 0.0    & 13.50 & 1.10 \\
 NGC 3960        & 1 & $-$0.15& 11.90 & 0.91 \\
 (8.9, 9.0, 9.1) & 2 & $-$0.1 & 12.00 & 0.95 \\
                 & 3 & $-$0.2 & 11.75 & 0.80 \\
                 & 4 & 0.0    & 12.10 & 0.75 \\

\hline

   \end{tabular}
 \end{center}
\end{table}


\section{Results}
   \label{results}

   In this section we discuss several aspects of our results. We first examine the effects of the starting value of the MCMC chain on the final posterior distribution. We then report our final, best-fit values for the cluster-wide parameters, as well as assess the BASE-9 fit by generating isochrones with the best-fit cluster parameters and plot these on the cluster CMDs. We also explore the effect of the prior distribution on our results, and then discuss asymmetric posterior distributions. Finally, we comment on the advantages of using BASE-9 for fitting cluster CMDs over traditional by-eye fitting methods, such as we presented in Section \ref{isochrones}.

\subsection{Starting Values}
   \label{sect_starting}

     For each cluster we did multiple runs and for each run we start the MCMC chain in a different location of parameter space, some of which are statistically distant from the target posterior distribution. If the chains return to the same distribution after a sufficiently long run, we conclude that the results are insensitive to our choice of starting value (this is based on the convergence diagnostic for MCMC chains based on multiple runs, as described by Gelman \& Rubin 1992). Thus, consistent results for all starting values is evidence that the MCMC technique is efficiently sampling the target posterior distribution. We perform these convergence tests in two ways: first, changing the starting value of age while using the same starting values of metallicity, distance, and reddening; and second, by varying the starting values of metallicity, distance, and reddening. 

\begin{figure*}
\begin{center}
    \epsscale{1.00}
     \plottwo{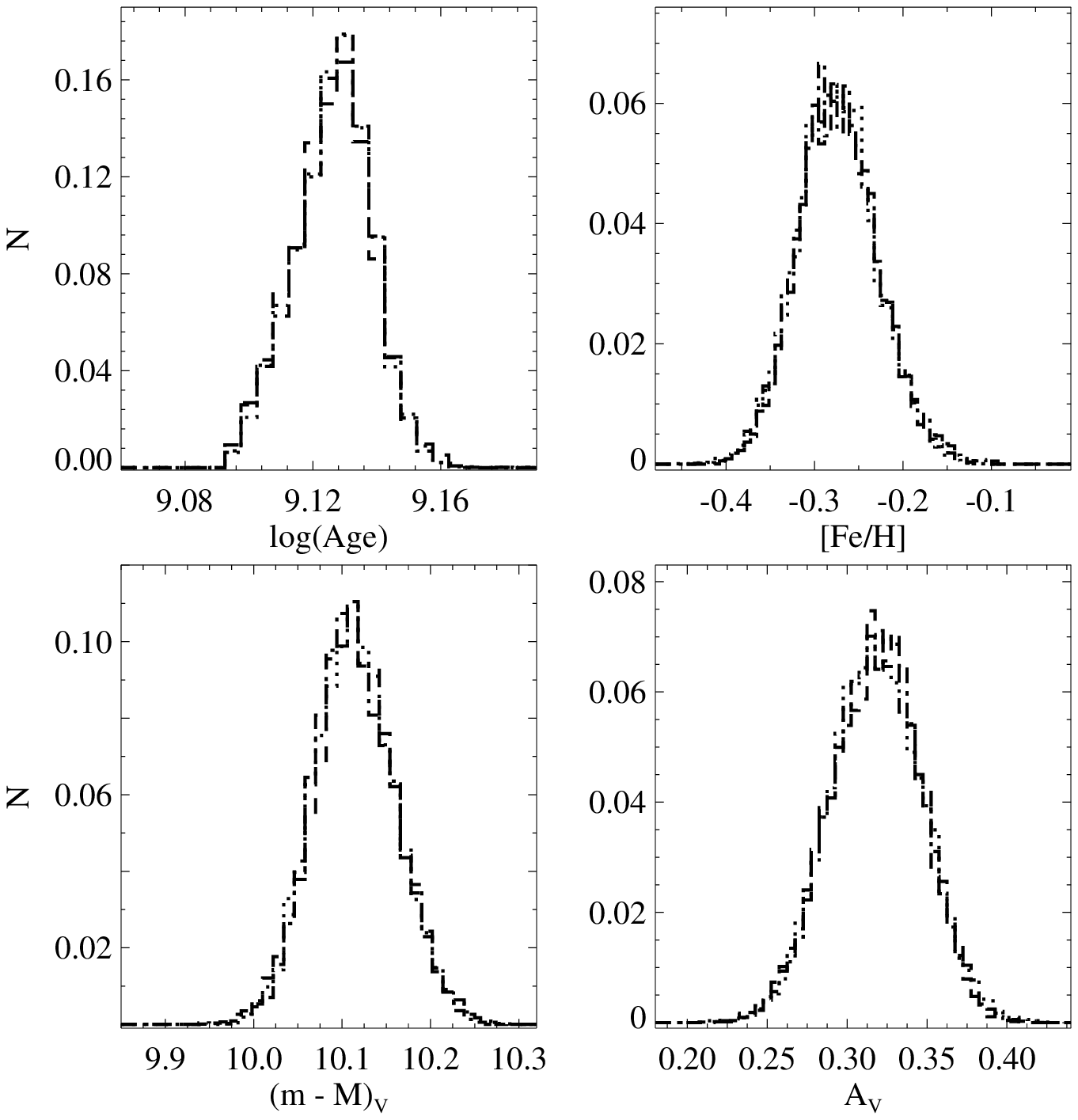}{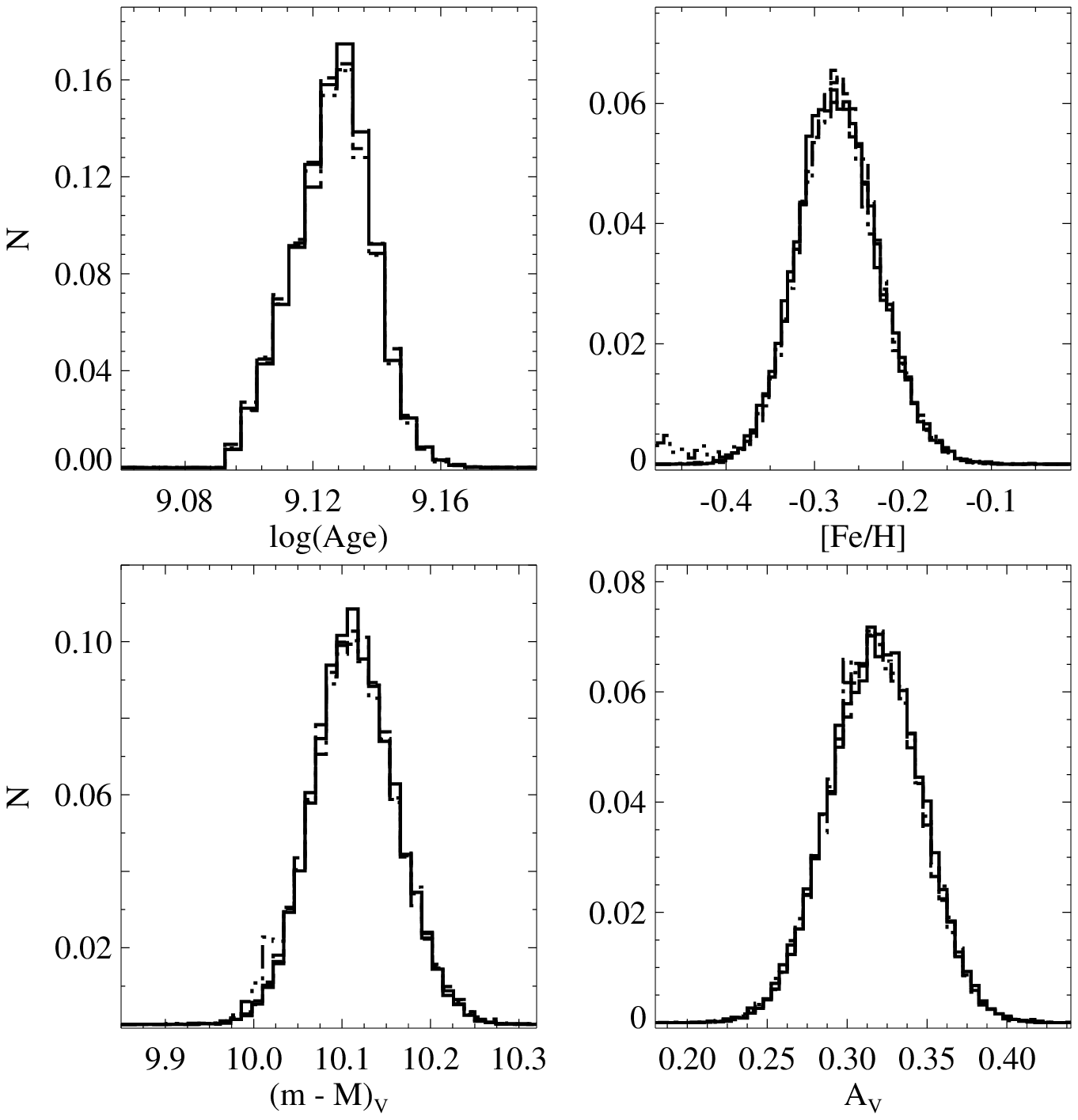}
	\caption{Left:   Posterior distributions of the three 
	runs of Set \# 1 for NGC 2360. Different runs are represented by 
	different line styles. The same starting values for metallicity, 
	distance, and reddening are used, but different starting values on age.
	Right: Posterior distributions for the four sets of 
	starting values for NGC 2360. All runs for a given set were combined 
	for comparison. Each set is indicated with a different line style. 
	Despite different starting values on all cluster parameters, BASE-9 
	consistently recovered the same posterior distribution, demonstrating 
	the robustness of the technique to starting values.}
\label{ngc2360_prior1}
\end{center}
\end{figure*}

   We use the case of NGC 2360 to illustrate these results (Figure \ref{ngc2360_prior1}). In the left panel of Figure \ref{ngc2360_prior1} we plot the posterior distributions (as histograms) of the four cluster parameters recovered by the three runs of Set \#1 for NGC 2360 (see Table \ref{starting}). Each run is represented by a different line style. As can be seen in this figure, despite different starting values of age, BASE-9 determined consistent solutions for each of the four cluster parameters. In the right side of Figure \ref{ngc2360_prior1} we plot the posterior distributions for the four cluster parameters for each of the four sets for NGC 2360 (Table \ref{starting}), again using a different line style to represent each set. Again, BASE-9 consistently found the same posterior distributions. Results of this test were the same among all sets for each of the clusters. From these convergence tests, we conclude that the Bayesian technique is robust in finding the posterior distributions, regardless of the starting value of age, metallicity, distance, or reddening.

\subsection{Best-fit Cluster Parameters}
   \label{best_fit}

\begin{figure*}[t]
\begin{center}
    \epsscale{0.9}
     \plotone{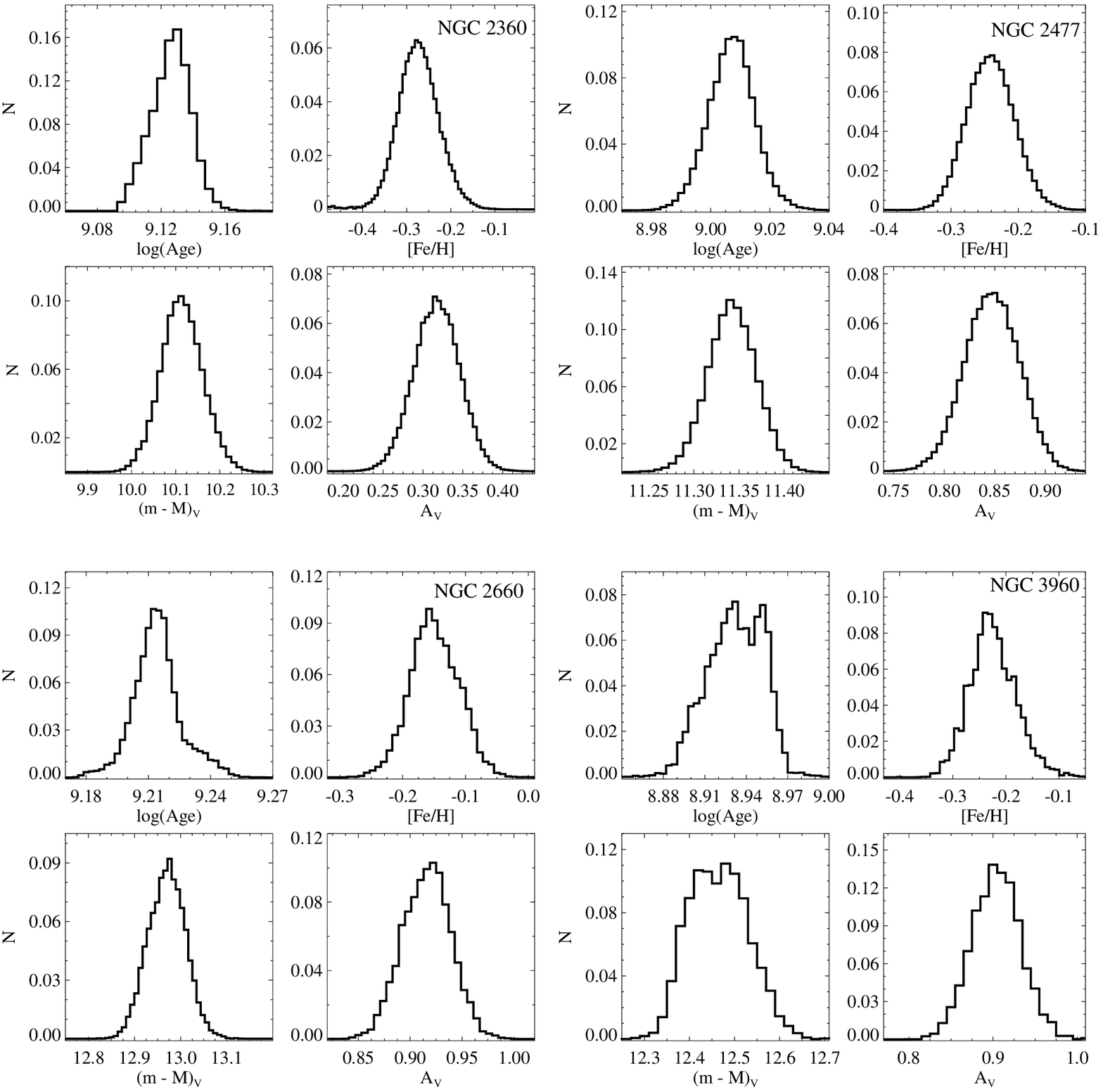}
         \caption{All posterior distributions for all 
        parameters of each cluster. The means and standard 
	deviations of these distributions are listed in Table 
	\ref{results_table}. Some distributions are noticeably 
	asymmetric (e.g., [Fe/H] in NGC 2660) and others are bimodal 
	(e.g., log(age) in NGC 3960).}
\label{all_pdfs}
\end{center}
\end{figure*}

   Given that our sample of the posterior distribution is independent of starting values, we combine all 12 MCMC chains for each cluster into a single posterior distribution for each cluster-wide parameter. We present these full posterior distributions for each cluster in Figure \ref{all_pdfs}.

   The best summary of our analysis are the complete posterior distributions. Yet, for simplicity, we report the mean and standard deviation of the combined MCMC chains in Table \ref{results_table}. Again, we emphasize that the precision reported here is internal precision. 


\begin{table*}[bt]
  \begin{center}
   \caption{Summary statistics of cluster parameters}
      \label{results_table}
  \begin{tabular}{cccccc}

    \hline
 Cluster &  log(Age) & Age (Gyr) & $[Fe/H]$ & $(m - M)_{V}$ & $A_{V}$ \\

    \hline


 NGC 2360 &  9.129 $\pm$ 0.012 & 1.35  $\pm$ 0.04 & $-$0.27 $\pm$ 0.05 & 10.12 $\pm$ 0.05 & 0.32 $\pm$ 0.03 \\
 NGC 2477 &  9.008 $\pm$ 0.008 & 1.02  $\pm$ 0.02 & $-$0.24 $\pm$ 0.04 & 11.35 $\pm$ 0.03 & 0.85 $\pm$ 0.03 \\
 NGC 2660 &  9.216 $\pm$ 0.012 & 1.64  $\pm$ 0.04 & $-$0.14 $\pm$ 0.04 & 12.97 $\pm$ 0.04 & 0.92 $\pm$ 0.02 \\
 NGC 3960 &  8.935 $\pm$ 0.021 & 0.860 $\pm$ 0.04 & $-$0.22 $\pm$ 0.04 & 12.47 $\pm$ 0.07 & 0.91 $\pm$ 0.03 \\

\hline

   \end{tabular}
\\
 \end{center}
\end{table*}


\begin{figure*}
\begin{center}
    \epsscale{0.95}
     \plotone{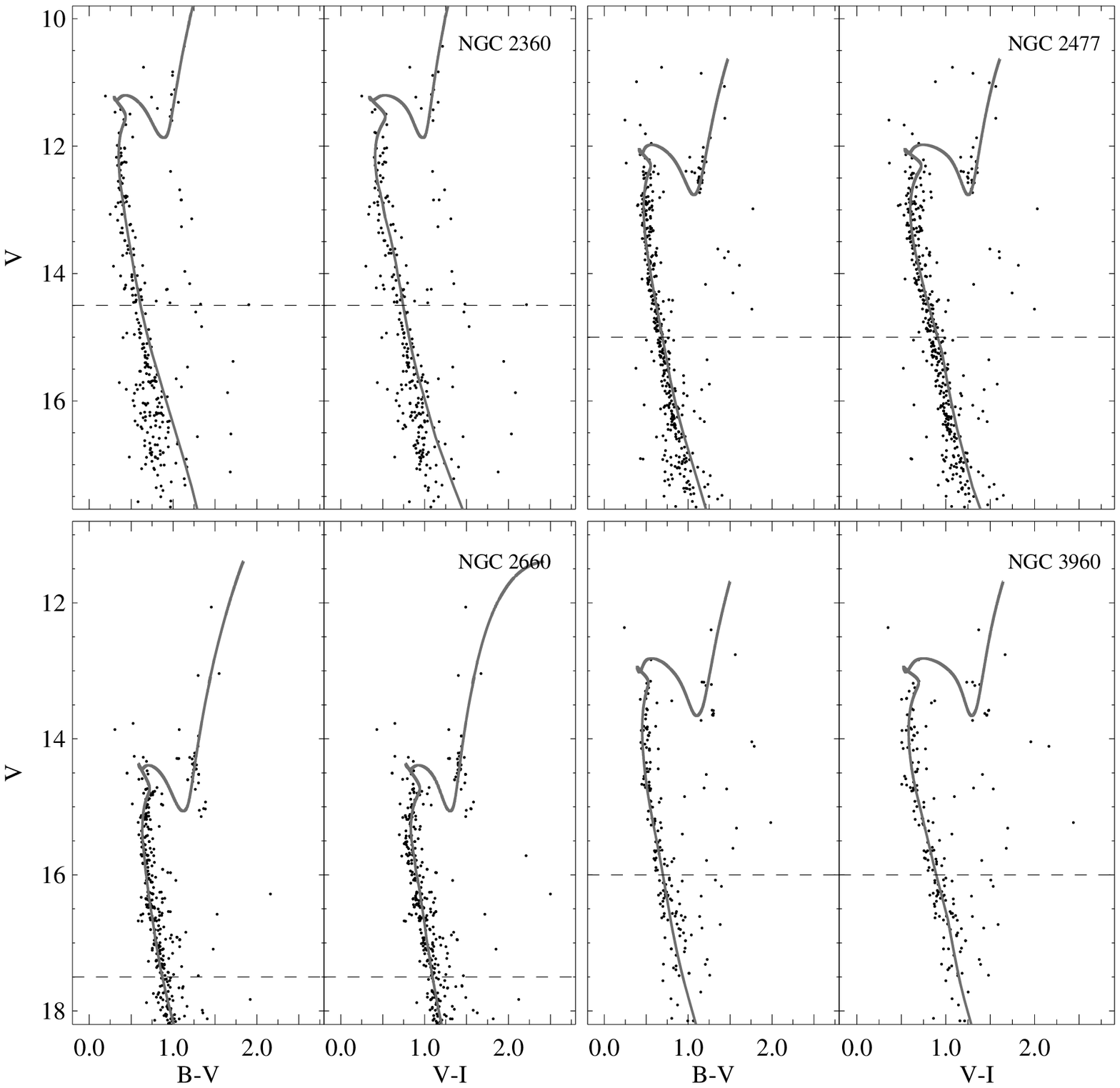}
         \caption{CMDs of each cluster with isochrones generated 
	using the best-fit parameters (Table \ref{results_table}), 
	as found by BASE-9. Only photometry used by BASE-9 are plotted.}
\label{cmd_bayes_iso}
\end{center}
\end{figure*}


   Although NGC 2477 and NGC 3960 have been shown to exhibit differential reddening (Hartwick et al. 1972; Bragaglia et al. 2006), our algorithm does not currently incorporate differential reddening. The small $\sigma$ values on $A_{V}$ should not be taken to imply that the clusters do not exhibit differential reddening.

   Using these best-fit cluster parameters, we generated isochrones to compare with the photometry. Doing so reinforces our confidence in the fit determined by BASE-9. In Figure \ref{cmd_bayes_iso} we present the CMDs with these BASE-9 determined isochrones. We retain the dashed horizontal line as a reminder of the magnitude limit employed by BASE-9. We note the excellent fit in every case.

\subsection{Dependence on Prior Distributions}
   \label{priors_dep}
   
      To investigate the dependence of our results on the prior distributions, we performed the following sensitivity tests. After obtaining the results discussed above, we again ran Set \#1 (see Table \ref{starting}) for each cluster four times with the following changes: (i) we doubled the prior $\sigma$ value on metallicity, leaving the other $\sigma$ values equal to the values in Table \ref{priors_table}; (ii) we doubled the prior $\sigma$ value on distance modulus, leaving the other $\sigma$ values equal to the values in Table \ref{priors_table}; (iii) we doubled the prior $\sigma$ value on reddening, leaving the other $\sigma$ values equal to the values in Table \ref{priors_table}; and (iv) we doubled the prior $\sigma$ values on metallicity, distance, and reddening. In every case the prior distribution on log(age) remained flat, as before.

   In all cases, results were similar to those discussed in the previous sections. Using NGC 2360 as an illustrative case, in Table \ref{prior_dep} we summarize the mean and standard deviation of each of the posterior distributions of the four runs described here, along with the original results of Set \#1 (using the original prior distributions listed in Table \ref{priors_table}) for comparison. As can be seen in this table, changing prior distributions caused most of the posterior distributions to shift less than one standard deviation from that of the original run. Results were similar for the other three clusters.
   
   From this we conclude that for these clusters and these data, sensible and even conservative variations on the prior distributions do not meaningfully influence the results. This increases our confidence in the posterior distributions obtained in Section \ref{best_fit}, especially for age, the parameter in which we are most interested. We note that changing the prior distribution on distance had the most notable effect on age; this sort of dependence will be mitigated when data that provides higher precision in the distances of clusters are available from Gaia.
   



\begin{table*}[bt]
  \begin{center}
   \caption{Prior Dependence Test Results for NGC 2360}
      \label{prior_dep}
  \begin{tabular}{ccccc}

    \hline
 Run &   log(Age) & [Fe/H] & $(m - M)_{V}$ & $A_{V}$ \\

    \hline


 Original                       &  9.129 $\pm$ 0.012 & $-$0.27 $\pm$ 0.05 & 10.12 $\pm$ 0.05 & 0.32 $\pm$ 0.03 \\
 $\sigma_{[Fe/H]} \times 2$     &  9.130 $\pm$ 0.012 & $-$0.29 $\pm$ 0.05 & 10.12 $\pm$ 0.05 & 0.33 $\pm$ 0.03 \\
 $\sigma_{(m-M)_{V}} \times 2$     &  9.135 $\pm$ 0.012 & $-$0.27 $\pm$ 0.04 & 10.09 $\pm$ 0.05 & 0.31 $\pm$ 0.03 \\
 $\sigma_{A_{V}} \times 2$         &  9.125 $\pm$ 0.012 & $-$0.32 $\pm$ 0.05 & 10.15 $\pm$ 0.05 & 0.36 $\pm$ 0.04 \\
 all $\sigma \times 2$          &  9.132 $\pm$ 0.012 & $-$0.35 $\pm$ 0.05 & 10.12 $\pm$ 0.05 & 0.37 $\pm$ 0.04 \\

\hline

   \end{tabular}
\\
 \end{center}
\end{table*}


\subsection{Complex Posterior Distributions}
   \label{nonGauss}

   One of the advantages of using a disciplined Bayesian method is the recovery of posterior distributions that may be asymmetric or even multi-modal. These types of distributions can lead to an increased understanding in, e.g., how individual stars can drive the solution. To illustrate this, we explore the bimodal posterior distribution of the age of NGC 3960 (see the lower right panels of Figure \ref{all_pdfs}).
   
\begin{figure*}
\begin{center}
    \epsscale{0.63}
    \plotone{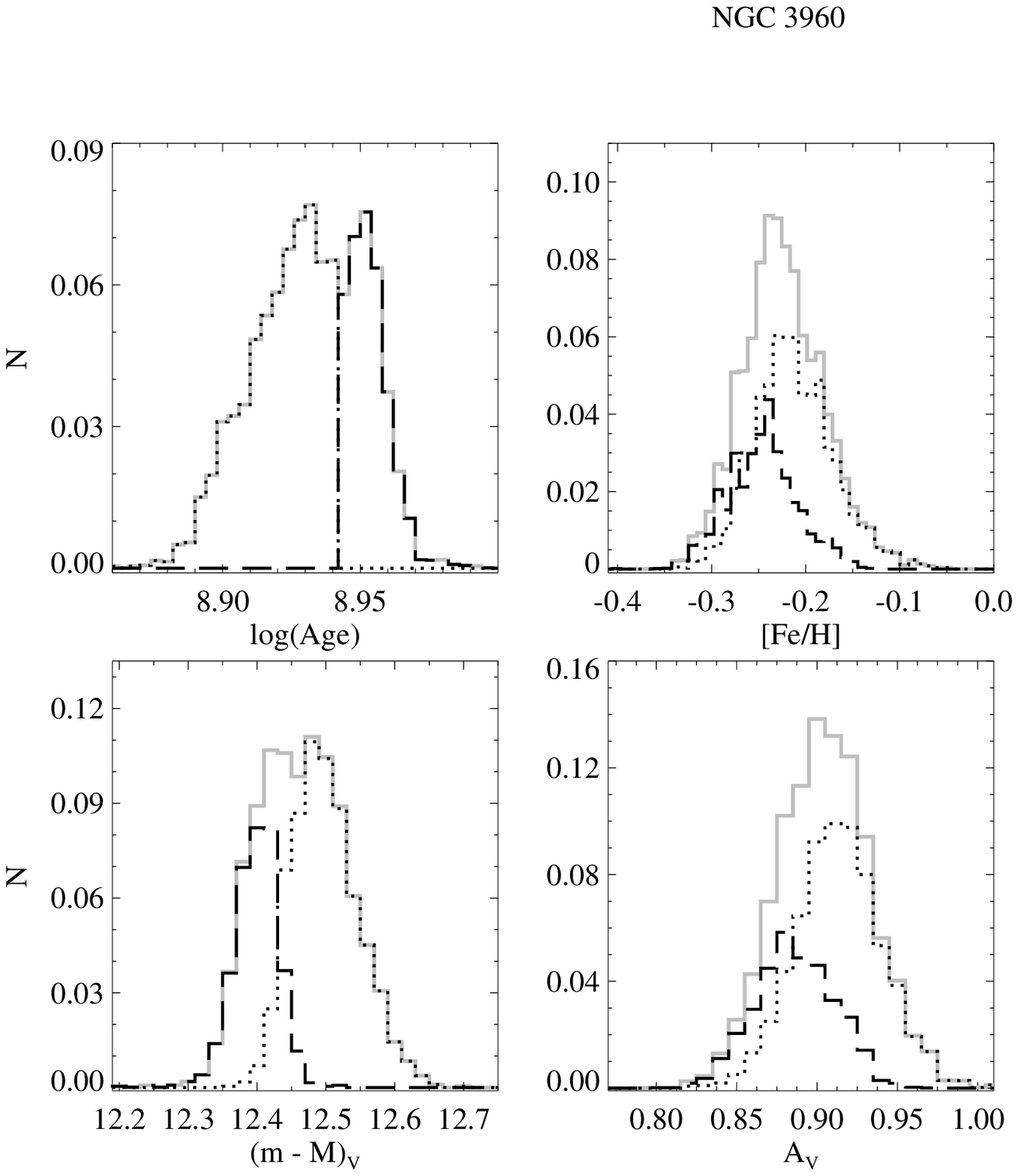}
	\caption{Original distributions for NGC 3960 (solid gray); a hard cut was done to separate the two 
	modes in the age distribution\---represented as the dashed and dotted lines. The draws from each part 
	of the distribution are then plotted using the same line style for metallicity, distance, and reddening. Note 
	that this also reproduces the bimodal distribution on the distance modulus. For comparison, isochrones were 
	produced using the averages of each of the two distributions for each parameter (see Figure \ref{cmd2modes}).}
\label{bimodal3960}
\end{center}
\end{figure*}

   First, we did a cut of the MCMC result of age to separate the two modes, as we show in Figure \ref{bimodal3960}. In this figure, we plot the complete posterior distributions (from Figure \ref{all_pdfs}) in gray, the draws from the left mode with the dotted line and the draws from the right mode with the dashed lines. In the remaining three panels we plot metallicity, distance, and reddening, and we see that these draws separate from each other, e.g., also explaining the bimodality of the posterior distribution of distance. Based on these distributions, we were able to produce and compare isochrones generated with the means of each distribution and compare their fits.
   
\begin{figure*}
\begin{center}
    \epsscale{0.85}
    \plotone{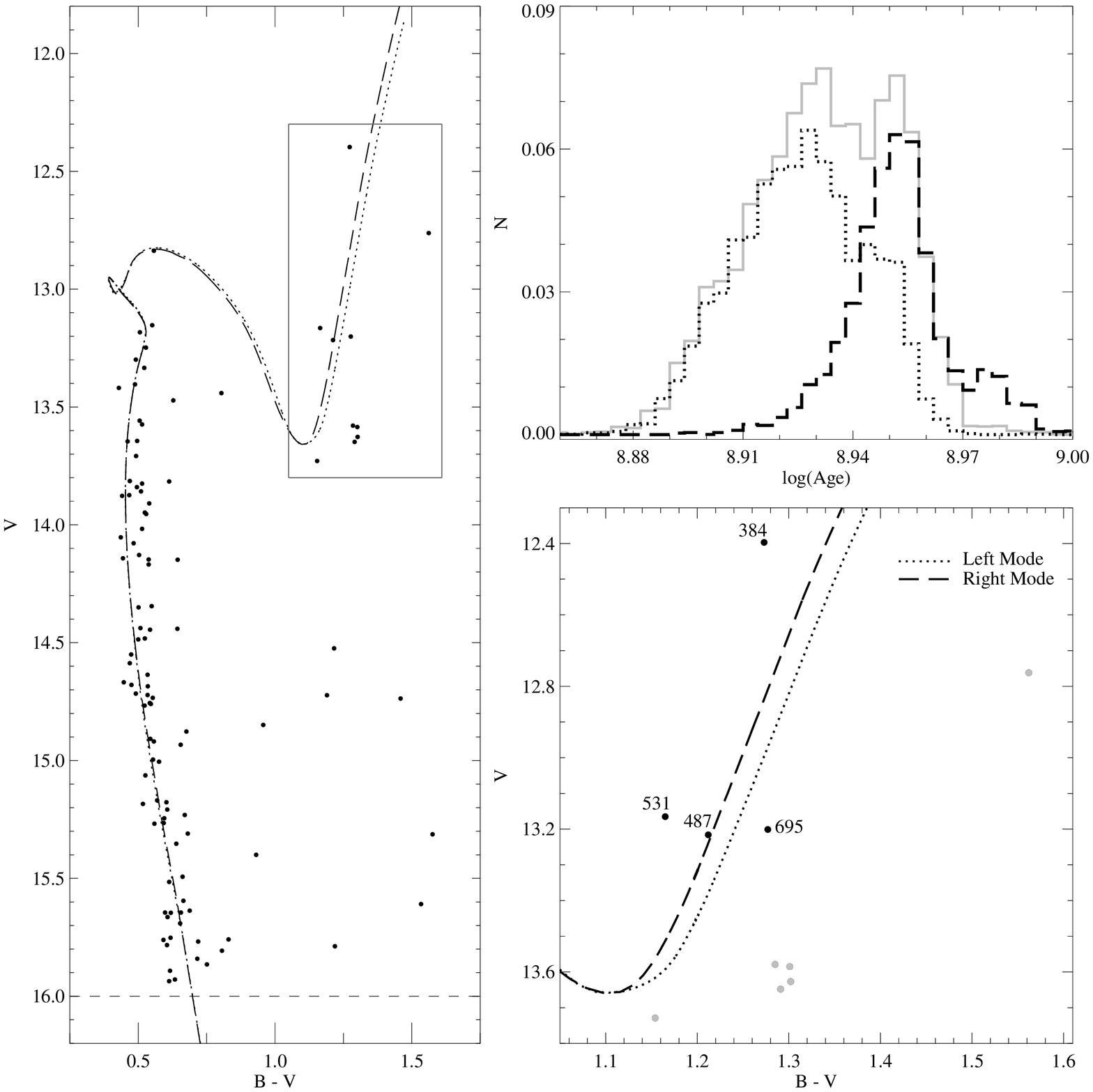}
	\caption{Left: $B-V$ CMD of NGC 3960 with isochrones overplotted. The isochrones 
	are generated to have cluster parameters corresponding the left and right mode of the age 
	distribution (see Figure \ref{bimodal3960}). The inset region is expanded in the lower right panel.
	Lower Right: Zoomed CMD; the four stars marked are those responsible for the bimodal age 
	distribution. The gray stars were consistently excluded as field stars.
	Upper Right: The original age distribution of NGC 3960 (solid gray), along with distributions 
	produced when different stars are considered cluster members or excluded from analysis. See the text for 
	full discussion.}
\label{cmd2modes}
\end{center}
\end{figure*}

   In the left panel of Figure \ref{cmd2modes} we overplot these two isochrones on the $B-V$ CMD of NGC 3960. As before, the dotted line represents the isochrone produced using mean values from the draws from the left mode of the age distribution, while the dashed line uses mean values for the right mode. The isochrone fits are very close but a small visible difference can be seen in the red giant branch. The gray box on the full CMD shows the region that is zoomed in the lower right panel of Figure \ref{cmd2modes}. In the  zoomed CMD, the gray stars are those that were consistently rejected as field stars by BASE-9. We investigated the effect of the four labeled stars (384, 487, 531, and 695) on the solution.
   
   We first re-ran the cluster using BASE-9, but this time we remove star 695 completely and set the prior probability of cluster membership of stars 384, 487, and 531 to 1.0. This forces those stars to be cluster members by not allowing BASE-9 to consider the possibility that they may be field stars. We then ran BASE-9 again, this time setting the prior probability of cluster membership of star 695 to 1.0 and removing stars 384, 487, and 531.
   
   The posterior distributions resulting from these tests are shown in the upper right panel of Figure \ref{cmd2modes}. The solid gray line shows the original age distribution for NGC 3960, with the other distributions overplotted and re-scaled arbitrarily for comparison. Based on these plots, the explanation for the bimodal age distribution is clear. BASE-9 identifies two possibilities for Stars 384, 487, 531, and 695: either star 695 is a cluster star and the others are not, or stars 384, 487, and 531 are cluster stars and star 695 is not. The first possibility corresponds to the left mode in the age distribution and the second possibility corresponds to the right mode. Looking at the zoomed CMD (bottom right panel) this is not surprising, as these stars straddle the isochrones, corresponding to the left and right modes.
   
   This test demonstrates the power of BASE-9 in isolating and understanding the role individual stars on the CMD can play, and understanding non-Gaussian distributions. We note that because the peaks of the two modes of the age distribution of NGC 3960 are within one standard deviation of the average of the total distribution (see Table \ref{results_table}), we retain the estimates and errors we previously reported for this cluster.

\subsection{BASE-9 vs. Traditional Fitting}
   \label{base_v_trad}

   In Section \ref{isochrones} we employed the long-used, eye-based technique of fitting isochrones to cluster CMDs to obtain ages, as well as other cluster-wide parameters.  Performing a best fit by eye can be difficult, as the effects of some parameters can mimic others in the CMDs. As can be seen from Figure \ref{cmd_iso}, the fits look good, but at times the crudeness of the technique limits confidence that the fits are optimum. 

   The by-eye technique is further complicated when photometry in multiple filters is available. The data used here are photometry in three filters ($BVI$), meaning we could have three possible CMDs for each cluster (i.e., $V$ vs. $B-V$, $V-I$, or $B-I$; more CMDs are possible by also varying the color on the vertical axis). We want to optimize over all of the CMDs. Plotted CMDs are, in reality, two-dimensional (2D) projections of what is really a higher dimensional diagram; various structures may not be apparent in these 2D projections. With BASE-9  we simultaneously fit isochrones to photometry in $all$ available filters, and these challenges disappear. The specific issue of fitting isochrones to a variety of filter combinations with BASE-9 has been explored in detail by Hills et al. (2015). They found that limitations in stellar models create systematic differences among some filter combinations, and find a general preference for the fits that involve more filters.

   The ages we determined using BASE-9 for NGC 2360 and NGC 2477 are within the uncertainty of the ages found using the by-eye technique (Table \ref{results_table}). For NGC 2660 and NGC 3960, the ages determined by BASE-9 are higher and lower (respectively) than those found using by-eye isochrone fitting, but fall among values found by previous authors (Tables \ref{lit2660} and \ref{lit3960}). Reasons for this discrepancy could include the difficulty encountered in fitting isochrones due to the abundance of field stars (of which these two clusters suffer from more than NGC 2360 or NGC 2477), or uncertainty in metallicity, distance, or reddening. In all cases, the error bars on age found using the Bayesian technique are considerably smaller, by an order of magnitude. More importantly, however, the most probable fit was determined in an objective and statistically robust way.

\section{Conclusion}
  \label{conclusion}

   We have employed a powerful software suite, BASE-9, to determine the best-fit isochrones for four intermediate-age open clusters: NGC 2360, NGC 2477, NGC 2660, and NGC 3960. Our primary interest is in high-precision cluster ages, which we determine to be 1.35 $\pm$ 0.05, 1.02 $\pm$ 0.02, 1.64 $\pm$ 0.04, and 0.860 $\pm$ 0.04 Gyr, respectively. This precision in age ranges from as little as 2\% to $<$ 5\% uncertainty. This approaches a new level in high-precision stellar cluster ages.

   Although by-eye methods can be used to approximate the best-fit values, they cannot achieve the high precision of principled Bayesian methods. Given the expense of quality modern data, it is important to use a robust statistical approach that maximally leverages these valuable data.

   We emphasize the importance of such an objective technique as a way to determine higher precision ages and cluster properties, making better use of and providing useful feedback for stellar evolution models.

\begin{acknowledgements}

   This material is based upon work supported by the National Aeronautics and Space Administration under grants No.\ NAG5-13070 and 10-ADAP10-0076 issued through the Office of Space Science, and by the National Science Foundation through grants AST-0307315, and DMS-12-09232. D.v.D. acknowledges support from a Wolfson Research Merit Award (WM110023) provided by the British Royal Society and from a Marie-Curie Career Integration (FP7-PEOPLE-2012-CIG-321865) and Marie-Skodowska-Curie RISE (H2020-MSCA-RISE-2015-691164) grants, both provided by the European Commission. D.C.S. acknowledges support from the European Research Council via an Advanced Grant under grant agreement No. 321323-NEOGAL. E.J.J. acknowledges support from the College of Physical and Mathematical Sciences and the Department of Physics and Astronomy at Brigham Young University.

\end{acknowledgements}

\vspace{0.2in}
\begin{center} \textbf{REFERENCES} \end{center}

\noindent
Bertin, E. \& Arnouts, S. A\&AS, 117, 393 \\
Bonatto, C. \& Bica, E. 2006, A\&A, 455, 931 \\
Bragaglia, A., Tosi, M., Carretta, E., et al. 2006, MNRAS, 366, 1493 \\
Bragaglia, A., Sestito, P., Villanova, S., et al. 2008, A\&A, 480, 79 \\
Carraro, G. 2014, 
Carraro, C., Ng, Y.K., \& Portinari, L. 1998, MNRAS, 296, 1045 \\
Claria, J.J., Piatti, A.E., Mermilliod, J.-C., Palma, T. 2008, AN, 329, 609 \\
DeGennaro, S. von Hippel, T., Jefferys, W.H., et al. 2009, ApJ, 696, 12 \\
Dotter, A., Chaboyer, B., Jevremovic, D, et al. 2008, ApJS, 178, 89 \\
Eigenbrod, A. Mermilliod, J.-C., Claria, J.J., et al. 2004, A\&A, 423, 189 \\
Friel, E.D \& Janes, K.A. 1993, A\&A, 267, 75 \\
Friel, E.D., Janes, K.A., Tavarez, M., et al. 2002, AJ, 124, 2693 \\
Geisler, D., Claria, J.J. \& Minniti, D. 1992, AJ, 104, 1892 \\
Gelman, A., \& Rubin, D.B. 1992, Statistical Science, 7, 457 \\
Gunes, O., Karatas, Y. \& Bonatto, C. 2012, New Astronomy, 17, 720 \\
Hamdani, S., North, P., Mowlavi, N., et al. 2000, A\&A, 360, 509 \\
Hartwick, F.D.A., Hesser, J.E. \& McClure, R.D. 1972, ApJ, 174, 557 \\
Hartwick, F.D.A. \& Hesser, J.E. 1973, ApJ, 183, 883 \\
Heiter, U., Soubiran, C., Netopil, M., et al. 2014, A\&A, 561, 93 \\
Hills, S. von Hippel, T., Courteau, S., et al. 2015, AJ, 149, 94 \\
Kassis, M., Janes, K.A., Friel, E.D., \& Phelps, R.L. 1997, AJ, 113, 1723 \\
Janes, K.A. 1981, AJ, 86, 1210 \\
Jeffery, E.J. von Hippel, T., DeGennaro, S., et al. 2011, ApJ, 730, 35 \\
Landolt, A.U. 1992, AJ, 104, 340 \\
Mazzei, P. \& Pigatto, L. 1988, A\&A, 193, 148 \\
Mermilliod, J.-C. \& Mayor, 1990, A\&A, 237, 61 \\
Mestel, L. 1952, MNRAS, 112, 583 \\
Meynet, G., Mermilliod, J.-C. \& Maeder, A. 1993, A\&AS, 98, 447 \\
Patenaude, M. 1978, A\&A, 66, 225 \\
Perryman, M.A.C., Brown, A.G.A., Lebreton, Y., et al. 1998, A\&A, 331, 81 \\
Prisinzano, L., Micela, G., Sciortino, S., et al. 2004, A\&A, 417, 945 \\
Reddy, A.B.S., Giridhar, S. \& Lambert, D.L. 2012, ASInC, 4, 197 \\
Salaris, M., Weiss, A. \& Percival, S.M. 2004, A\&A, 414, 163 \\
Sandrelli, S., Bragaglia, A., Tosi, A., et al. 1999, MNRAS, 309, 739 \\
Sarajedini, A., von Hippel, T., Kozhurina-Platais, V. \& Demarque, P. 1999, AJ, 118, 2894 \\
Sestito, P., Bragaglia, A., Randich, S., et al. 2006, A\&A, 458, 121 \\
Stein, N.M., van Dyk, D.A., von Hippel, T., et al. 2013, Statistical Analysis and Data Mining, 6, 1, 34 \\
Stenning, D.C., Wagner-Kaiser, R., Robinson, E., et al. 2016, ApJ, 826, 41 \\ 
Taylor, B.J. \& Joner, M.D. 2005, ApJS, 159, 100 \\
Twarog, B.A., Ashman, K.M. \& Anthony-Twarog, B.J. 1997, AJ, 114, 2556 \\
van Dyk, D.A. et al. 2009, The Annals of Applied Statistics, 3, 117 \\
VandenBerg, D.A., Bergbusch, P.A. \& Dowler, P.D. 2006, ApJS, 162, 375 \\
von Hippel, T., Gilmore, G., Jones, D.H.P. 1995, MNRAS, 273, 39 \\
von Hippel, T., Robinso, E., Jeffery, E., et al. 2014, arXiv:1411.3786 \\
von Hippel, T., Jefferys, W.H., Scott, J., et al. 2006, ApJ, 645, 1436 \\
Winget, D.E., Hansen, C.J., Liebert, J., et al. 1987, ApJ, 315, L77 \\

\end{document}